\documentclass[fleqn,twoside]{article}

\usepackage{espcrc2}
\usepackage{graphicx}
\usepackage{amssymb}

\def\Nf{N_{\rm f}}
\def\csw{c_{\rm SW}}
\def\kc{\kappa_{\rm c}}
\def\ksea{\kappa^{\rm sea}}

\def\mps{m_{\rm PS}}
\def\mv{m_{\rm V}}
\def\Qtop{Q_{\rm top}}

\title{%
       \vspace{-1.88cm}					
       {\normalsize DESY 01--167} \\			
       \vspace{+1.40cm}					
       Low-lying fermion modes of $\Nf =2$ improved Wilson fermions%
       \thanks{Talk given by D.~Pleiter at the Lattice Conference 2001,
               Berlin, Germany.}
}

\author{%
  {\it QCDSF} and {\it UKQCD} Collaboration:
  R.~Horsley\address[Ztn]{John von Neumann Institute NIC / DESY Zeuthen,
                            D-15738 Zeuthen, Germany},
  T.G.~Kov\'{a}cs\addressmark[Ztn],
  V.~Linke\address[FU]{Institut f\"{u}r Theoretische Physik,
                       Freie Universit\"{a}t Berlin,
                       D-14195 Berlin},
  D.~Pleiter\addressmark[Ztn],
  G.~Schierholz\addressmark[Ztn]$^{,}$%
               \address[HH]{Deutsches Elektronen-Synchrotron
                            DESY, D-22603 Hamburg, Germany}
}

\begin{document}

\begin{abstract}
We present preliminary results for the topological charge and susceptibility
determined from the low-lying eigenmodes of the Wilson-Dirac operator.
These modes have been computed on dynamical configurations with $\Nf = 2$
non-perturbatively improved Wilson fermions. We compare our results with
the eigenmodes of fermions in the quenched approximation.
\end{abstract}

\maketitle

\section{INTRODUCTION}

For several reasons there is particular interest in the low-lying eigenmodes
of the Dirac operator in QCD.
From phenomenological models it is expected that for sufficiently light
quark masses physics, e.g. hadron correlators, is dominated by these modes.
Calculations on the lattice enable us to study the relevance of
low-lying eigenmodes for various observables directly.

Furthermore, the eigenmodes of the Dirac operator carry information about
the topological content of the background gauge field.
The topological properties can also be probed using gluonic methods. These
however typically suffer from large fluctuations on very short scales
and therefore require some sort of smoothing procedures. Usually
the quantities of interest are affected by the applied filtering method.

Results from fermionic methods on the other hand are expected to
be sensitive to the
effects caused by explicitly breaking chiral symmetry when using Wilson
fermions. Although the symmetry is expected to be restored in the continuum
limit, these effects are potentially large on coarse lattices.
However, these can be reduced by applying Symanzik's improvement program.
Removing discretization effects of $O(a)$, where $a$ is the lattice
spacing, is achieved by adding a single counterterm to the action.
The improved Wilson-Dirac operator then reads:
\begin{equation}
M = 1 - \kappa\,\left[H + \frac{i}{2}\,\csw F_{\mu\nu}\sigma_{\mu\nu}\right],
\label{eq:dirac}
\end{equation}
where $H$ is the hopping term of the usual Wilson fermion action.

For the calculations presented here we used O(20-50) dynamical configurations
with $\Nf=2$ flavours of degenerate quarks for six different combinations
of $\beta$ and $\ksea$.
We used lattices of size $16^3\times 32$ with a lattice spacing $a$ that
varies between 0.104(1) and 0.095(1) fm (using the force scale $r_0=0.5$ fm).
For the lightest and heaviest sea
quarks the ratio $\mps/\mv$ was 0.60 and 0.83, respectively. For comparison
we performed the same calculations using quenched configurations at
$\beta=6.0$ and $a = 0.093(1)$ fm.

\section{COMPUTING LOW-LYING EIGENVALUES}

\begin{figure}[th]
\vspace*{0.12cm}
\begin{center}
\includegraphics[scale=0.30]{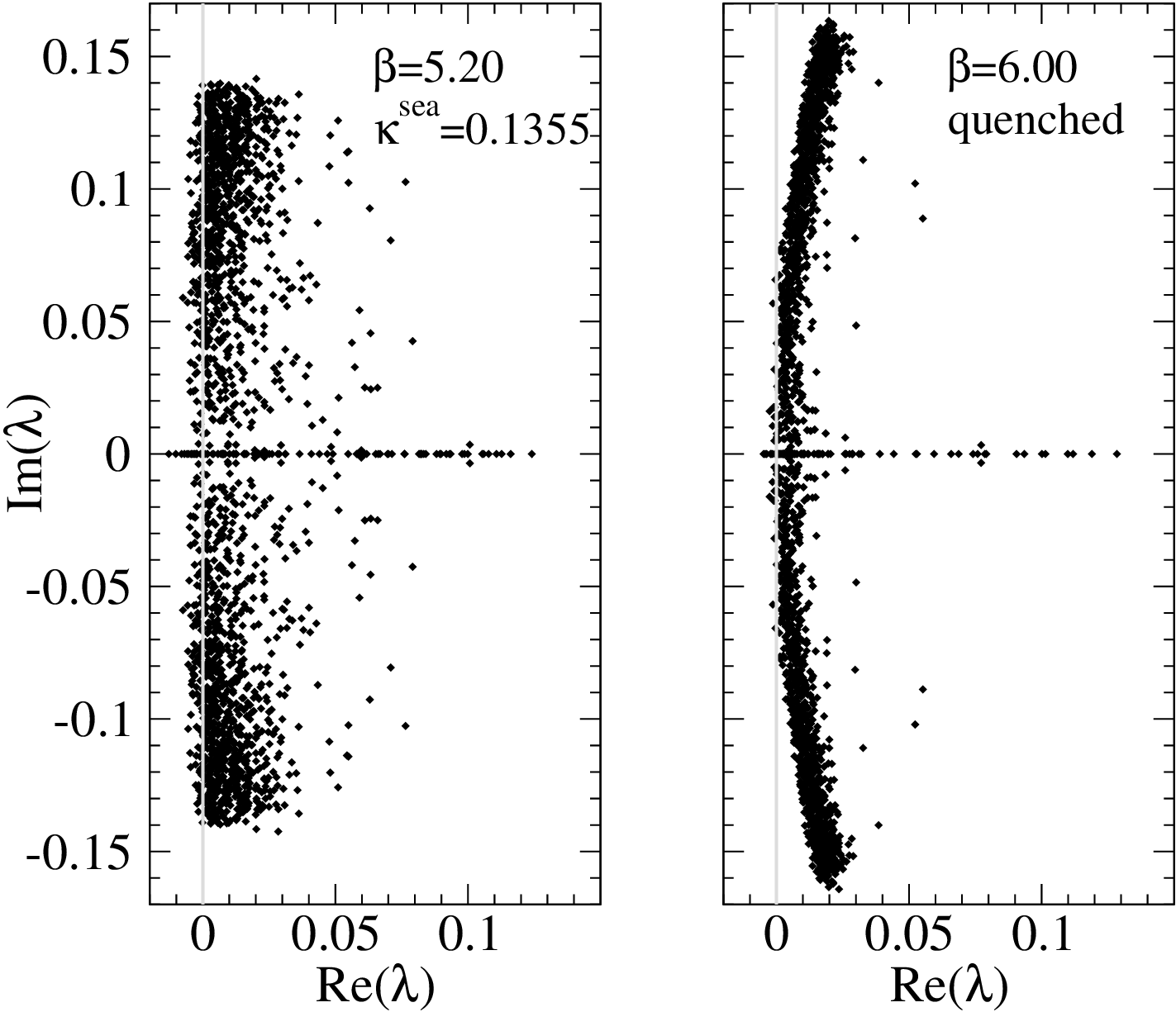}
\end{center}
\vspace*{-1cm}
\caption{\label{fig:spectrum}%
O(100) smallest eigenvalues of the massless Dirac operator calculated
on 20 gauge dynamical (left) and quenched (right) configurations.}
\vspace*{-0.5cm}
\end{figure}

The Wilson-Dirac operator $M$ is a non-hermitian operator.
The chirality $\omega_i$ of an eigenmode $\lambda_i$
can be calculated using the corresponding right eigenvector $v_i$:
\begin{equation}
\omega_i = \sum_t \omega_i(t) =
      \sum_t\sum_x v^{\dagger}_i(x,t)\,\gamma_5\,v_i(x,t).
\end{equation}
We can distinguish between two types of eigenmodes. Firstly,
complex modes with non-vanishing imaginary part and $\omega_i = 0$ and,
secondly, real modes with vanishing imaginary part and $\omega_i \neq 0$.

In principle, it is possible to find the real modes by determining
the zero modes of the hermitian operator $\gamma_5 M$. If
$\lambda = 1 - \kappa\sigma$ is a real mode of $M$, then $\gamma_5 M$
will have a zero mode for $\kappa = 1/\sigma$. In practice it might
however be difficult to detect several real modes which lie close
to each other.  We therefore decided to calculate the 70-100 smallest
eigenvalues of the non-hermitian operator $M$ using the Arnoldi
algorithm~\cite{arpack}. Results for the massless Dirac operator
\begin{equation}
D = \frac{1}{2\kappa} M -
    \frac{1}{2}\left(\frac{1}{\kc} - \frac{1}{\kappa}\right)
\end{equation}
are plotted in Fig.~\ref{fig:spectrum}.

\section{REAL MODES}

\begin{figure}[t]
\vspace*{0.12cm}
\begin{center}
\includegraphics[scale=0.30]{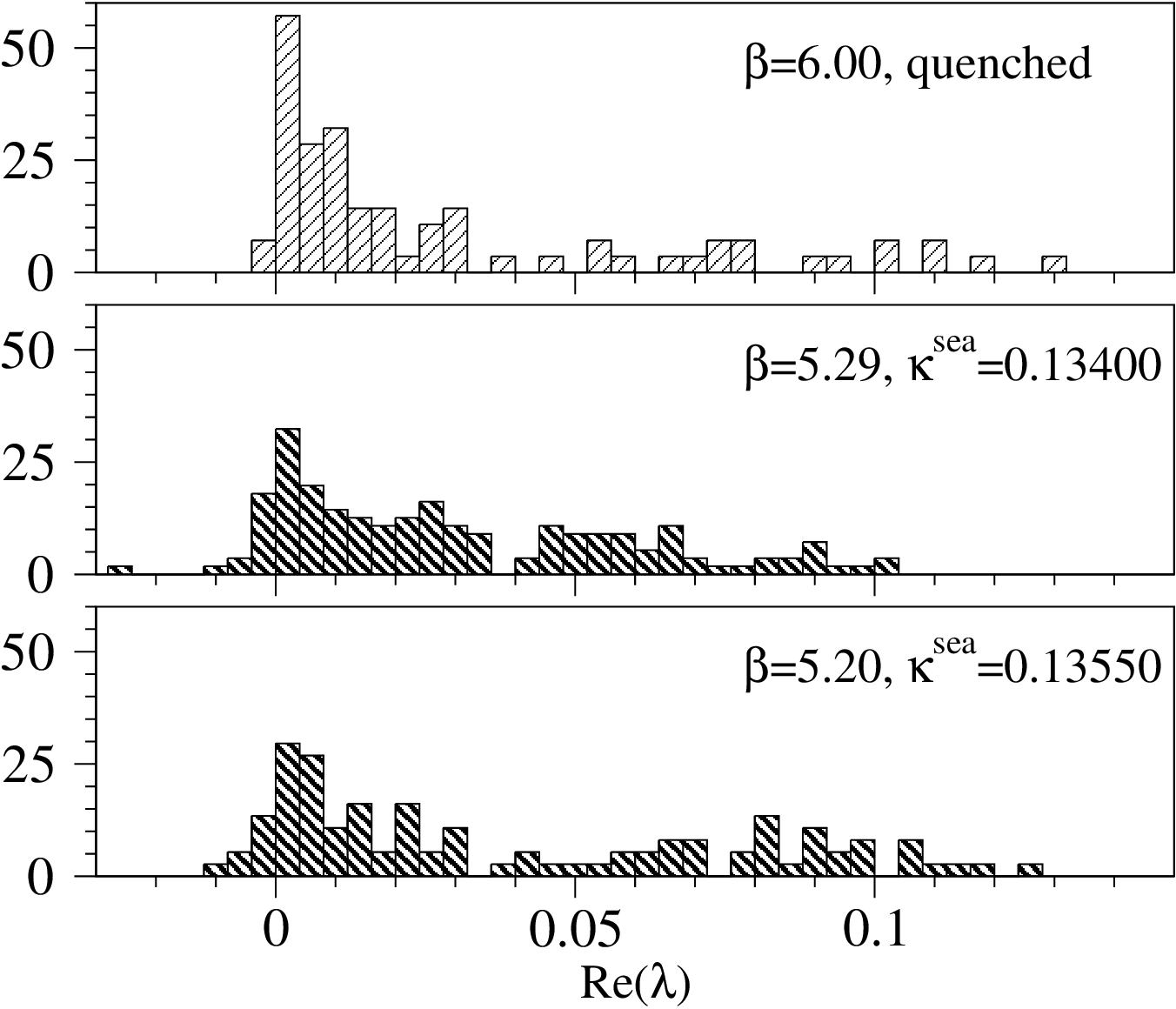}
\end{center}
\vspace*{-1.3cm}
\caption{\label{fig:densreal}%
Density of the real modes as a function of the real part. The area
under the curve is normalized to one. The graphs are ordered from
heaviest, i.e. infinite, (top) to smallest (bottom) sea quark mass.}
\vspace*{-0.3cm}
\end{figure}

For dynamical fermions one would expect eigenmodes around zero to be
suppressed. As can be seen from Fig.~\ref{fig:spectrum}, this seems
indeed to be the case. This effect is more pronounced if one looks
at the density of the real modes (Fig.~\ref{fig:densreal}).

Real modes close to (or below) zero, which correspond to zero modes
of $\gamma_5 M$ at $\kappa \lesssim \kc$, are believed to be the origin of
so-called exceptional configurations, which are encountered frequently
in the quenched approximation when approaching the chiral regime.
Although these modes are expected to be suppressed for dynamical fermions,
there is a non-zero probability for this kind of configurations to occur
(see Fig.~\ref{fig:densreal}).

\begin{figure}[t]
\vspace*{0.12cm}
\begin{center}
\includegraphics[scale=0.30]{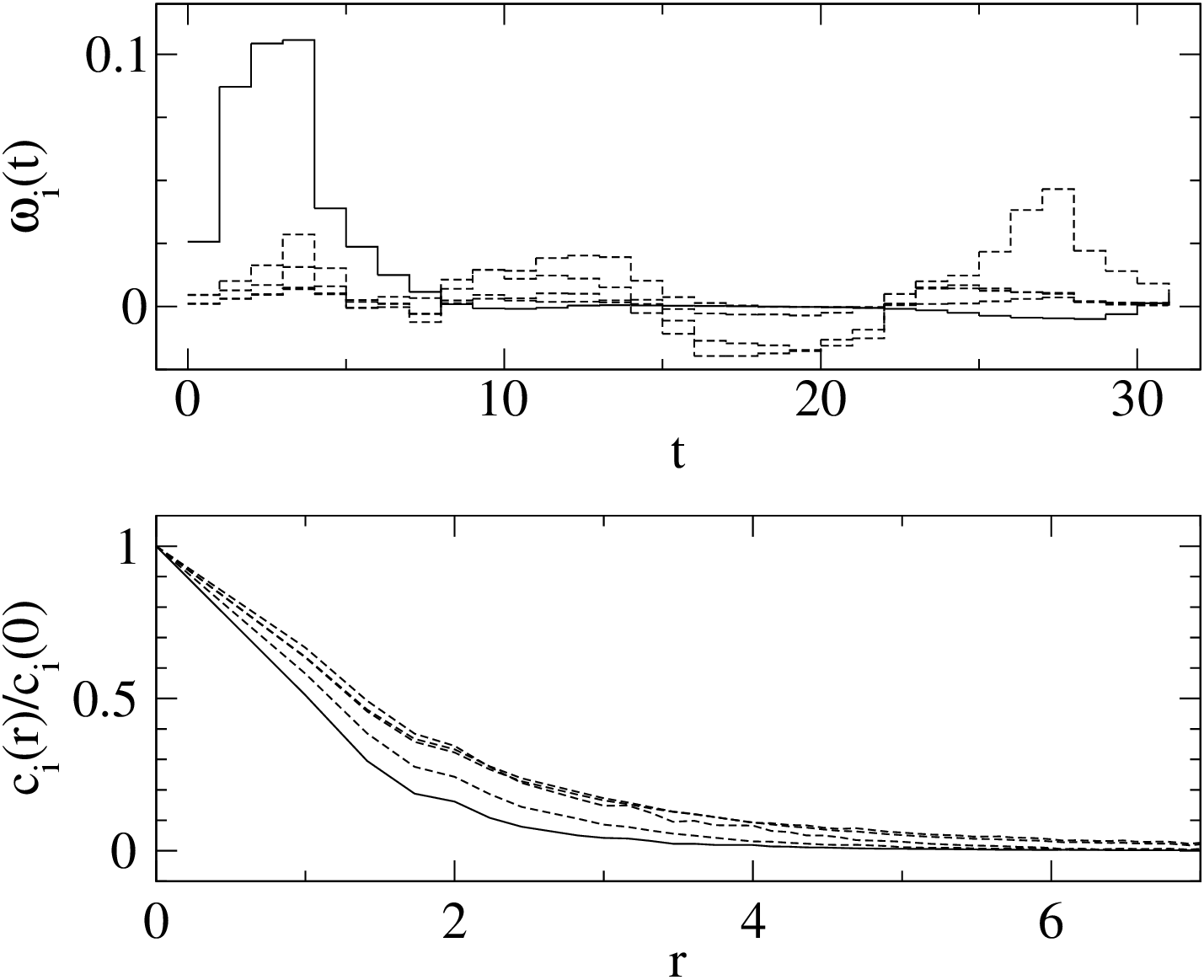}
\end{center}
\vspace*{-1.2cm}
\caption{\label{fig:corr}%
Chirality per timeslice $\omega_i(t)$ for real modes (upper picture)
and the normalized correlation function of the chirality (lower
picture) $c_i(r)$ on a configuration at $\beta=5.29$, $\ksea=0.13450$. The
real mode with the smallest eigenvalue is plotted with a solid line.}
\vspace*{-0.5cm}
\end{figure}

The correlation of the chirality
$c_{i}(r) = (1/V) \sum_{x,t} \omega_i(x,t) \omega_i(x+r,t)$ shown in
Fig.~\ref{fig:corr} was found to be narrow,
confirming the observations made in~\cite{DeGrand:2001gq}.
Typically the chirality of small real modes is concentrated in a few
time slices.

\section{TOPOLOGICAL SUSCEPTIBILITY}

The topological susceptibility normalised by the volume is defined by
\begin{equation}
\chi = \frac{\langle \Qtop^2\rangle}{V}.
\end{equation}
The topological charge $\Qtop$ can be obtained via
the Atiyah-Singer index theorem, which is approximately true on the lattice.
It establishes a relation between the
difference in the number of real modes with negative ($n^+$) and
positive ($n^-$) chirality and the topological charge:
\begin{equation}
\Qtop = n^+ - n^-
\end{equation}
For a subset of configurations we compared the results for $\Qtop$
with those obtained from gluonic operators using the Boulder
smoothing technique. The maximum difference found was $\pm 2$.
For a more detailed comparison see~\cite{alistair}.

\begin{figure}[th]
\begin{center}
\vspace*{0.1cm}
\includegraphics[scale=0.30]{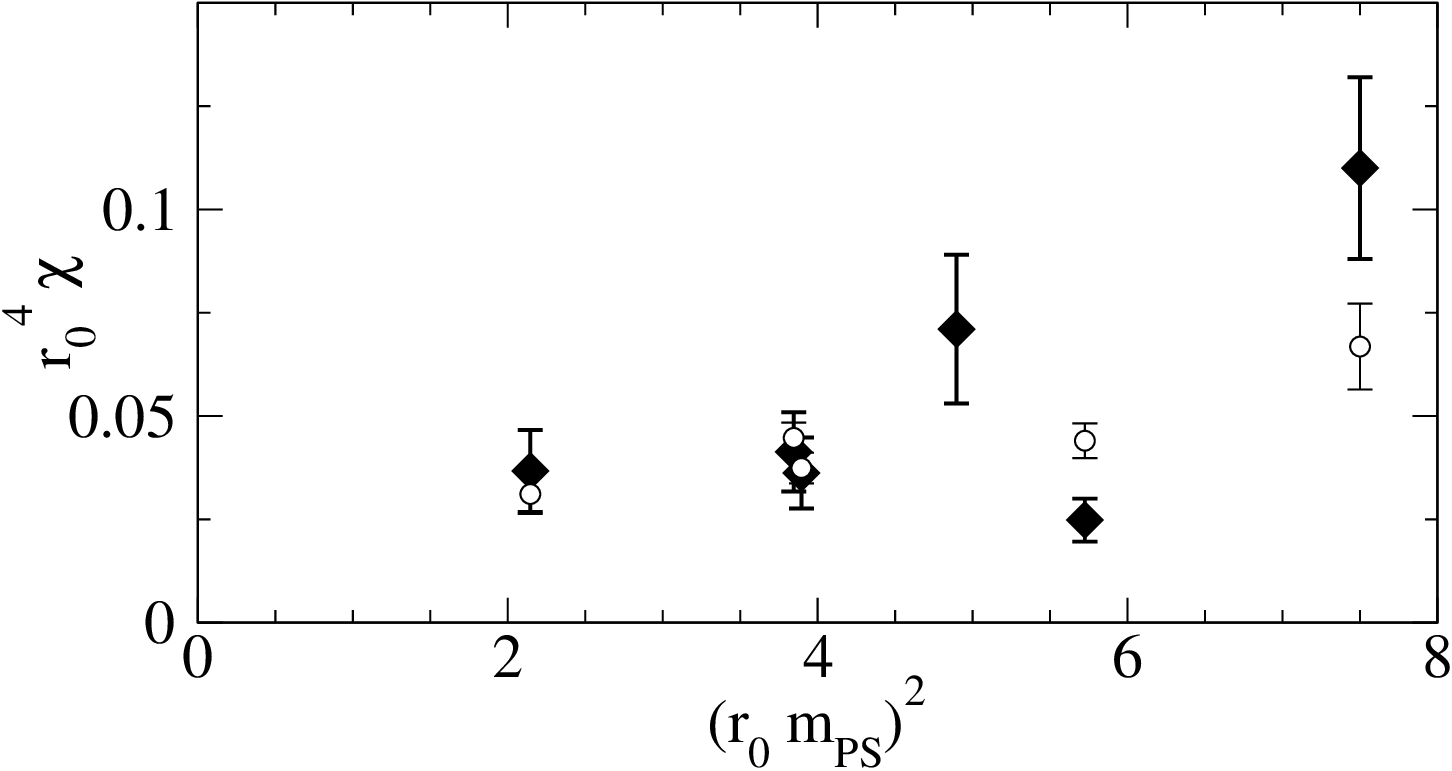}
\end{center}
\vspace*{-1.2cm}
\caption{\label{fig:suscept}%
Topological susceptibility as a function of the pseudoscalar meson mass.
The solid symbols show the results from this work, while UKQCD results
using gluonic methods~\cite{Hart:2001fp} are plotted with open symbols.}
\vspace*{-0.4cm}
\end{figure}

\begin{figure}[ht]
\begin{center}
\includegraphics[scale=0.30]{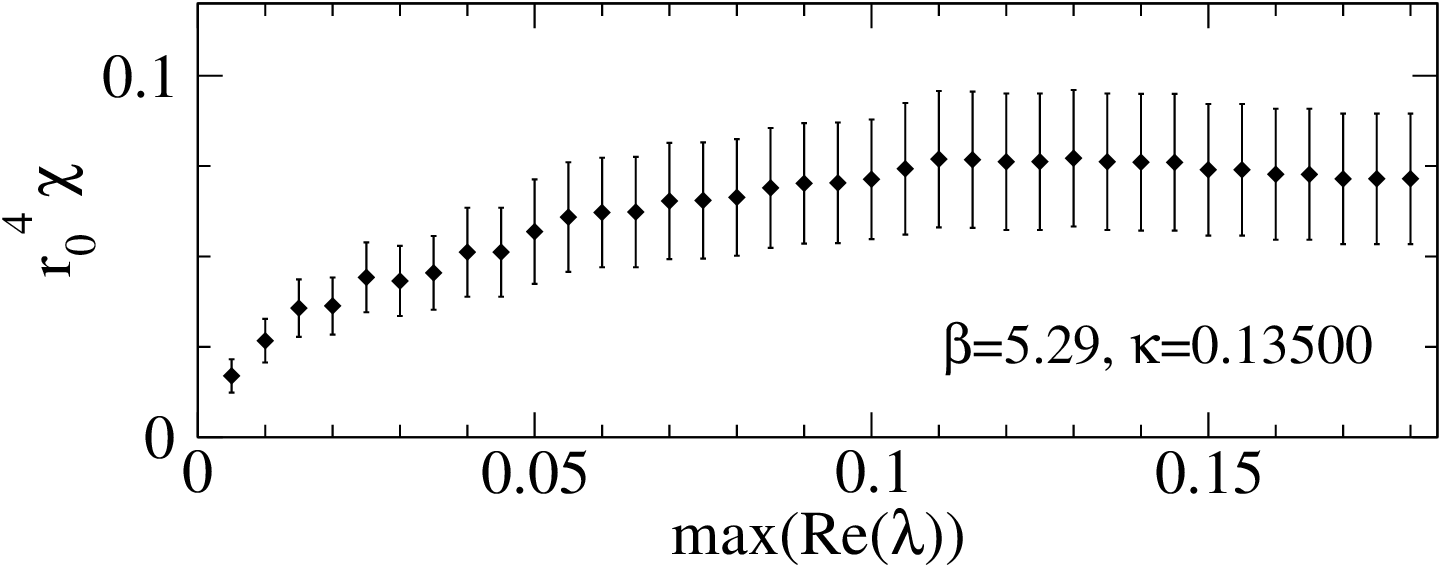}
\end{center}
\vspace*{-1cm}
\caption{\label{fig:plateau}%
Topological susceptibility versus the cut-off of used real modes.}
\vspace*{-0.5cm}
\end{figure}

The results for the topological susceptibility are plotted in
Fig.~\ref{fig:suscept}.
The discrepancy between the results from the fermionic method applied
in this work and from the gluonic method used by Hart and Teper is
likely to be explained by a lack of statistics. Another source of
uncertainty arises from the fact that we only calculated the lowest
70-100 eigenvalues. Although the probability for real modes is expected to
decrease when moving away from zero, it does not become zero. To
check for this possible error, we looked at the topological susceptibility
as a function of the largest real mode taken into account
(see Fig.~\ref{fig:plateau}). For all data sets we found a reasonable
plateau.

\section{CONCLUSIONS}

Comparing real modes of quenched and dynamical configurations we
found evidence for an expected depopulation of the region around
zero. The results for $\Qtop$ obtained by using the Atiya-Singer-theorem
were in reasonable agreement with numbers obtained from gluonic methods.
Our preliminary results for the topological susceptibility were found
to be similar to other calculations using the same configurations
but gluonic methods.

\section*{ACKNOWLEDGEMENTS}

The numerical calculations were performed on the Hitachi {SR8000} at
LRZ (Munich), the Cray {T3E} at EPCC (Edinburgh) and ZIB (Berlin).
We wish to thank all institutions for their support.

\end{document}